# Opto-thermal transport engineering in hybrid organic-inorganic lead halide perovskites metasurfaces

Saswata Halder[†], Amit Kessel, Noa Mazurski and Uriel Levy[†]

Department of Applied Physics, Faculty of Science and the Center for Nanoscience and Nanotechnology, The Hebrew University of Jerusalem, Jerusalem, 919041, Israel.

**ABSTRACT:** Halide perovskites have recently gained widespread attention for their exceptional optoelectronic properties which have been illuminated by extensive spectroscopic investigations. In this article, nanophotonic surface-engineering using soft-lithography has been used to reproduce nanostructures with enhanced functionalities. A non-invasive optical technique based on Raman and photoluminescence (PL) spectroscopy is employed to investigate the interactive effect of the thermal and optical behaviour in surface-patterned hybrid organic-inorganic halide perovskite thin films. The thermophysical properties of the engineered perovskite films are extracted from the softening of the representative peak positions in the Raman and PL spectra of the samples which act as temperature markers. The investigation suggests a comparatively higher rise in the local temperature for the patterned thin films resulting from their enhanced absorption. Therefore, a cross-talk between the opto-thermal transport phenomena in imprinted perovskite thin films pertaining to both enhancing device properties along with maintaining device stability is established.

The hybrid organic-inorganic halide perovskite $CH_3NH_3PbI_3$ (MAPbI) with an $ABX_3$ architecture has recently been demonstrated as successful optical and optoelectronic material due to its direct band gap, long carrier lifetimes, high electronic and hole mobilities and diffusion lengths [7, 8]. The low cost and facile fabrication techniques of MAPbI thin films in solution form are advantageous in terms of their large scale industrial generation [9-11]. Nano-patterning of the perovskite thin films have been shown to improve the performance in solar cells in terms of their efficiency through reduced surface reflectance, tailored light trapping, and higher external radiative efficiency and directionality [12-14]. In light emission applications, nanotexturing increases the out-coupling efficiency and imparts new features such as directionality and spectral tenability [15]. From the earliest techniques of fabrication by etching of the substrate and subsequent conformal coating to direct patterning of the perovskite film, researchers have demonstrated different approaches to create patterned perovskite metasurfaces. [16-19] Focused ion beam milling has been used to construct textures over small areas [20] but a scalable technique that takes advantage of the solution-based processing of the perovskite is more desirable. Toward this goal, nanoimprint lithography has been applied to pattern perovskite films through heated and pressurized contact with inflexible stamps.[21-23] Direct patterning of solution-deposited hybrid perovskites using flexible polydimethylsiloxane (PDMS) stamps was also demonstrated. [24-26] Unlike a silicon stamp, the PDMS can imprint its textures conformally on non-uniform surfaces and over defects. [27] The process is also recyclable enabling the usage of a single master to reproduce the same pattern on different

PDMS stamps ensuring the method to be cost effective. To ensure optimal optoelectronic functionalities in the visible region, the feature size on the stamp must be pushed to the nano-regime, which however requires attention to the integrity of the stamp. Features as small as a few nanometers have been replicated in molded polymers such as polyurethane and in silica sol–gel films.[27-29] Imprinting nanoscale features into crystalline materials, such as crystalline polymers, however, is far more challenging because interfaces with the stamp and substrate dramatically influence the crystallization process.[30] These nanophotonic textures can potentially be used to enhance the efficiency of perovskite solar cells, direct the emission of light emitting diodes, or generate perovskite distributed feedback lasers (DFB) lasers. Such imprinting techniques are particularly appropriate for halide perovskites because their sensitivity to water and plasma processing makes them incompatible with conventional lithography and etching. In order to develop high quality nanoscale imprinting processes, it is essential to control the crystallization dynamics of the perovskite materials.

Lead halide perovskites have been known to be particularly sensitive to exposures to air, humidity and heat. Almost at the same time, many researchers found the instability of perovskites despite their excellent electronic properties.[31, 32] As far as we are concerned, perovskites are found to be sensitive to moisture, oxygen, UV light, light soaking, heat, electric field, and other potential factors.[32-35] The thermal stability and light soaking are the most challenging concerns affecting PSCs stability, since it is hard to avoid temperature increase and light illumination for solar cells during operation. One of the contributing factors for the heat sensitivity of the perovskites is the ultralow thermal conductivity of the material. The low thermal conductivity (0.2-0.5 $WK^{-1}m^{-1}$) is a result of the dynamics of the organic (MAI) component of the perovskite arising from the rotations of the MA cations as well as the overall tetragonal-to-cubic phase transformations [36-41]. The thermal conductivity in MAPbI originates from phonon transport resulting from the anharmonically coupled interaction between the organic cation and inorganic cage. The Raman spectra of MAPbI perovskites have been studied in details by many researchers [42-48]. MAPbI perovskite structure feature a significant dynamical disorder which is aptly visualised through lattice vibrations (phonons) and molecular rotations [49-51]. Whereas electron-phonon interactions have a substantial influence on the solar cell operations, the phonon-phonon scattering process is directly related to the thermal properties of the perovskite [52-54]. The low thermal conductivity in perovskites are directly related to the phonon lifetimes. The phonon lifetime is essential as it provides critical insights into the different electron-phonon and phonon-phonon couplings occurring in the material [55]. Ab-initio molecular dynamics (AIMD) simulations along with third order perturbation theory have predicted very short phonon lifetimes ~ 20 ps and ~ 1 ps at the Brillouin zone center and Brillouin zone boundaries respectively, consistent with the findings from experimental inelastic scattering processes with X-rays and neutrons [56-59]. Such a short phonon lifetime means that the phonons do not efficiently dissipate heat and can therefore affect the carrier relaxation and phonon-phonon scattering processes. The inefficient dissipation of heat can result in localised increase in temperature accelerating the destabilisation process for the halide perovskites.

As the lattice dynamics of MAPbI perovskites can provide an idea about the coordination environment as well as thermal properties of the material [56, 60-61] one of the easiest and non-destructive ways to probe its optical as well as thermal properties simultaneously is by using the Raman and PL spectroscopic techniques [62-64]. The laser beam effectively serves as: (i) a source of photons which are scattered inelastically by the material, and (ii) a heat source which increases the local temperature of the material at the point of incidence through absorption. The increase in the local temperature depends on the thermal conductivity of the material, therefore temperature and excitation power dependent optical measurements enable us to map the thermal properties of the materials to the lattice vibrations. In this paper, we use Raman spectroscopy to demonstrate the cross-talk between the optical and thermal transport mechanism of surface patterned MAPbI perovskite thin films. We adopt a surface imprinting method coherent with a modified micro-imprinting of a thin spin-coated precursor MAPbI perovskite film in soft-gel state with a topographically prepatterned elastomeric PDMS mould yielding large area crystalline structures. The analysis of the Raman active modes as a function of temperature and laser power for both the un-patterned and patterned perovskites are performed to extract the change in the local thermal behaviour of materials. Temperature dependent PL measurements are initiated to correlate the temperature sensitivity of the optical properties in the nanostructured films. The article concludes with a possible estimation of the thermal conductivity of the planar and patterned perovskite films, showing the use of differential spatial patterning to create thermal gradients for futuristic thermoelectric materials.

Fig. 1(a) demonstrates schematically the idea of probing differential thermal gradients besides photoluminescence enhancement in textured perovskite films developed by soft lithographic techniques. The perovskite thin film is characterized with the help of X-ray diffraction technique as shown in Fig. 1(b). The thin film show characteristic peaks of MAPbI with no secondary phases arising from degradation by products ($PbI_2$) confirming the single phase of the material [66]. The crystal structure of MAPbI is shown in Fig 1(c). The core-level XPS spectra for MAPbI is shown in Fig. S1 of the supplementary information showing the characteristic signature peaks [67]. The Pb 4f spectrum exhibits peaks at 138.9 eV and 143.8 eV corresponding to the Pb $4f_{7/2}$ and Pb $4f_{5/2}$ state associated to the $Pb^{2+}$ state in $PbI_3$ - octahedra. The C-1s peak shows two peaks corresponding to the presence of C-H/C-C and C-N bonds. The I-3d orbital shows the presence of two peaks at 619.7 eV and 631.2 eV corresponding to $I-3d_{5/2}$ and $I-3d_{3/2}$ states respectively. Fig. 1(d) show the SEM images of the stripe and nanohole patterned perovskite thin films. The structures are negative replications of the original PDMS mould. It is quite evident that the patterned samples have a smoother surface with only little porosity as compared to the pristine un-patterned film. The height of the grating is 140 nm, with a width and period of 1.3 μm and 2.6 μm respectively. The average diameter of the nanohole is 493 nm and the mean depth is 127 nm. These results demonstrate that although the perovskite is an ionic compound with no glass-transition behaviour, it can still be structured by soft lithographic methods and is soft enough to deform and be moulded into cavities under the application of external heat and pressure. The use of the elastomeric stamp has a favourable crystallization effect on the thin film as it reduces regions of lower surface energy and prevents unwanted nucleation of the crystals forming isolated large arrays with pores as evident from Fig. S2 of the supplementary information. This is also expected to have a positive effect on the electronic transport properties of the material as it provides more percolation pathways for the electron to diffuse upon the application of an electric field. Fig 1(e) shows the absorption spectra for the planar, stripe and nanohole patterned thin films. The nanoholes show higher absorption than the planar and stripes which paves the way for higher photon absorption in the visible region of the optical spectrum. The material also shows a high efficient PL spectrum at 765 nm with a full width half maximum of 43 nm besides showing a high refractive index. Therefore, it is possible to enhance the PL through the excitation of optical modes [68].

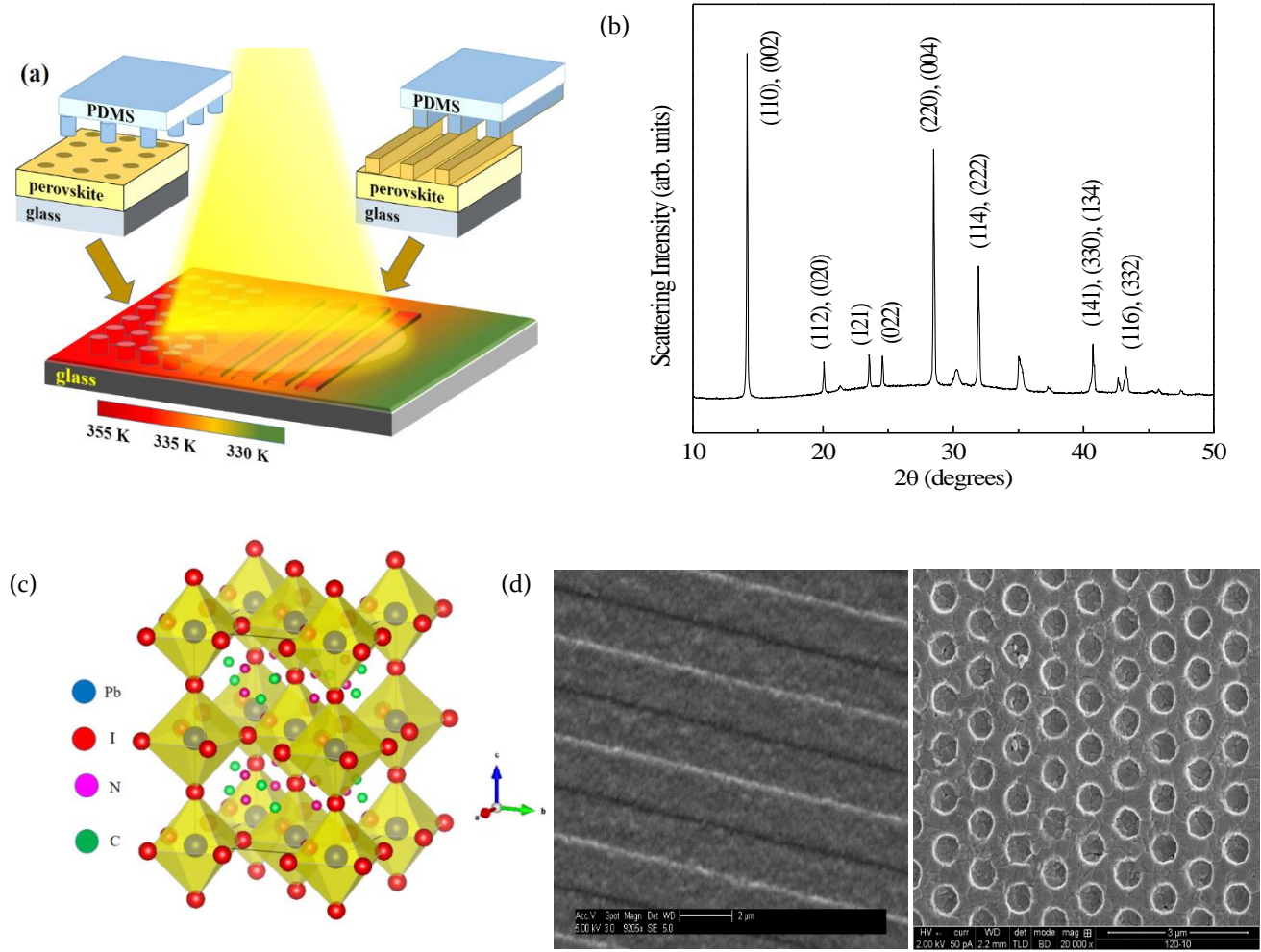

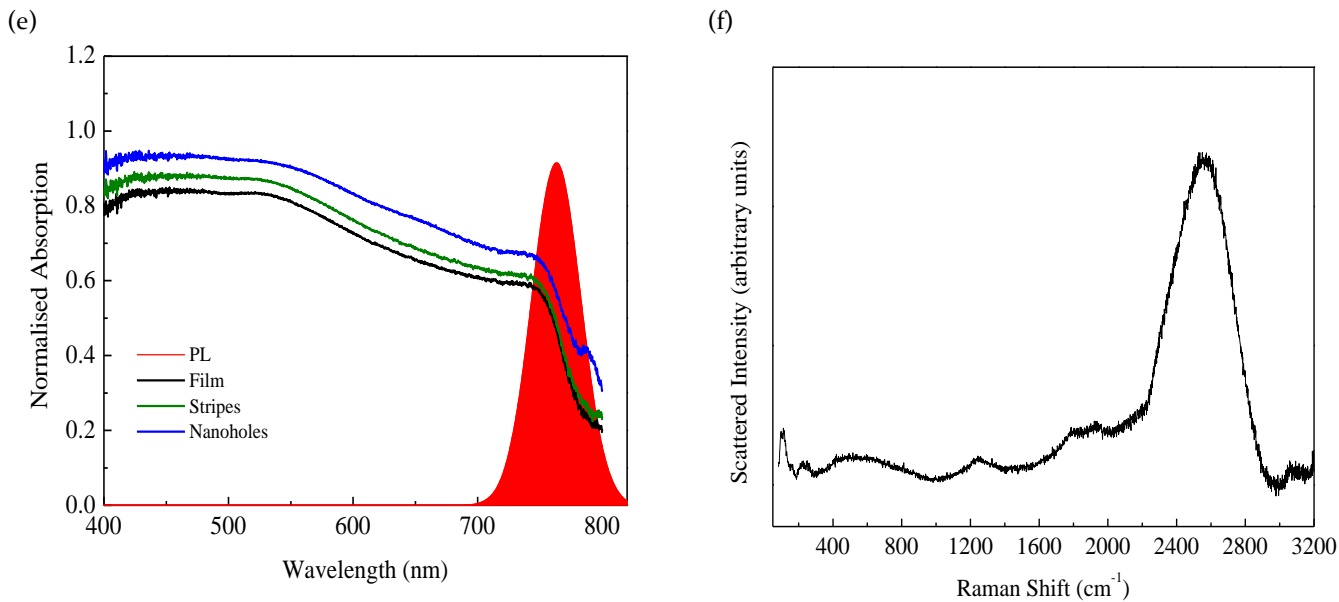

Figure 1: (a) Schematic representation of surface patterned perovskites and their corresponding properties (b) X-ray diffraction spectrum of MAPbI. The representative peaks are denoted by their corresponding (h,k,l) planes (c) The crystal structure of MAPbI. The blue, red, pink and green spheres represent lead (Pb), iodine (I), nitrogen (N) and carbon (C) respectively. (d) The scanning electron microscope (SEM) images of MAPbI stripes and nanoholes (e) The Raman spectra of MAPbI (f) The absorption and PL spectra of MAPbI pristine, stripes and nanoholes.

Fig. 1(f) shows the Raman spectra of a MAPbI film obtained under the laser excitation of 514 nm.

The spectrum is well characterized by broad curves at different wavelength intervals as shown in Figure S3 of the supplementary information. The peak positions can be compared to values reported in the literature for MAPbI [39, 51]. The structure of the Raman spectrum at lower wavenumbers (50–150 cm$^{-1}$) consists of a number of contributions at 98, 109, 130, 143 and 157 cm$^{-1}$. The spectrum at higher wavenumbers regime is dominated by the contributions from 214, 250, 340, 448 and 620 cm$^{-1}$. From a qualitative group theoretical analysis, the number of eigenmodes in this wavenumber domain should be 18, of which three are acoustic vibrations and the rest 15 are optical modes. However, due to the low strain in the system, most of the modes are energetically degenerate and therefore their experimental resolution is difficult. In the perovskite structure, the organic MA cation undergoes different reorientations about its lattice position besides coupling to the inorganic lead-halide framework, thereby further complicating the overall dynamical picture. The low wavenumber features are associated with cage dominated optical modes consisting of the overall translation and rotation of the organic cation coupled to the lattice motions. Inspection of the normal modes calculated in the region around 94 cm$^{-1}$ reveals a major component from librations of the organic cations and from Pb-I stretching, as shown in Figure 2b. Thus, the 94 cm$^{-1}$ band is reasonably associated with both the Pb−I stretching, I-Pb-I bending and to libration modes of the organic cations, and it mainly provides information regarding the inorganic component of the material. As pointed out previously, the fact that the Raman spectra calculated by the different models are quite different in the 100−150 cm$^{-1}$ region suggests that the band at 110 cm$^{-1}$ is probably associated with the motion of the organic cations. The band falling at 119 cm$^{-1}$ can be tentatively associated with the band calculated at 141 cm$^{-1}$ [69]. A similar result holds for the weak band measured at 157 cm$^{-1}$, that can be safely associated with the band calculated at 156 cm$^{-1}$, corresponding to the libration of the organic cations [69].

The region between 100 and 200 cm$^{-1}$ is associated to librational and molecular motions of the MA cations due to the mismatch in the mass of the organic and inorganic cation. Density functional theory (DFT) based electronic structure calculations elucidate how the interactions between the organic cations and the inorganic counterpart may affect the vibrational frequency and the Raman intensity of the MA torsional mode in MAPbI3 [51]. The broad and unresolved band at 200−340 cm$^{-1}$ is assigned to the torsional mode of the C-N bonds in the MA cations. The frequency of the MA torsional modes in the crystal structures may be the result of two competing effects, associated with the interactions between the cation and the inorganic cage. On the one hand, the deformation of the MA molecule shifts the torsional mode toward higher frequencies; on the other hand, the formation of specific hydrogen bonding interactions, typical of these compounds shifts this mode towards lower frequencies. The higher wavenumber peaks above 1000 cm$^{-1}$ result from $CH_3$ asymmetric breathing, $NH_3$ asymmetric breathing, C-H symmetric stretching and N-H symmetric stretching respectively [51].

In order to investigate the thermal conductivity of the material, temperature-dependent Raman spectra for the pristine perovskite along with the patterned perovskites are measured. The interaction between the phonon waves and the rotational degrees of freedom of the MA cations determine the attenuation of the thermal conductivity, resulting in the ultralow thermal conductivity of the material [70]. The deconvoluted temperature-dependent Raman profile for the three samples for 110 cm$^{-1}$ Raman shifts are displayed in Figs. 2(a)-(c) respectively. The main Lorentizian fitting is shown in Figs. S4-S6 respectively. The peak positions decrease with increase in temperature, which indicates that the thermal properties of the samples are modified due to application of heat [71-73]. The frequency shift due to phonon-phonon coupling is represented by [71, 73]:

$$\Delta\omega(T) = A\left[1 + \frac{2}{\exp\left(\frac{\hbar\omega_0}{2k_BT}\right)-1}\right] + \omega_0[\exp(-3\gamma\beta T) - 1] \quad (1)$$

here $\omega_0$ is the Raman frequency at 0 K, A is the anharmonic constant, $\gamma$ is the Gruneisen parameter and $\alpha$ is the coefficient of thermal expansion. From the ratio between the anti-Stokes and Stokes lines, the local temperature can be estimated. However, as anti-Stokes lines cannot be measured in the Raman set-up, the peak positions or full width at half maximum (FWHM) can be used as an appropriate temperature marker. The Raman shift in a sample is independent of factors such as resonant Raman, absorption, fluctuations in the laser pulses, etc. So, the uncertainties in estimating the temperature from the peak positions are lower than those from the peak width. The Raman shift as a function of temperature for bulk perovskite, perovskite grating and perovskite nanohole are displayed in Figs. 2(d)-(f), respectively. In both the striped and nanohole perovskite, the temperature reaches around 335 K and 355 K respectively whereas in bulk perovskite the temperature is lower and reaches a value of ~ 330 K. The local temperature rise depends on the heat transport properties of the materials. Intuitively, a local temperature increase will be lower for a material with higher thermal conductivity. A higher local rise in temperature is observed in patterned perovskites compared to pristine sample. Considering a three-photon scattering process, the scattering mechanism is empirically defined as $\tau^{-1} = \tau_b^{-1} + \tau_p^{-1} + \tau_u^{-1}$, where $\tau_b^{-1}$ is due to the scattering from sample boundaries, $\tau_p^{-1}$ is due to scattering from point defects like vacancies, substitutions and other point-like impurities. With increase in temperature, phonon-phonon interactions start to dominate as given by the three-phonon Umklapp scattering $\tau_u^{-1} \propto T^3$. $\tau_p^{-1}$ is mostly a bulk property of the sample and the effect of nanopatterning on it can be ignored as the patterning does not inherently change the vacancy-defect densities in the materials. Firstly, an increase in the absorption plays a role in increased energetic photon absorption. Due to the inherent ultralow thermal conductivity of the material, the spread of optically deposited heat will be hindered. Since, the number of photo excited carriers is low and their contribution to the thermal transport is negligible, the main contribution of the high energy incident photons will be to heat the material. Secondly due to altered grain and grain boundary dimensions arising as an artifact of the patterning process, boundary scattering also contributes to the thermal conductivity although their effect is much less than anticipated. Therefore the redshift in the phonon branches results from an increase in the phonon-phonon scattering. Thus, the phonon-phonon interactions in the nanopatterns may be more due to the increased rattling motion of the MA cations in the perovskite arising from the increased local heating of the sample. In a way, the metasurfaces act as 'localized heat concentrators', which concentrates the heat onto the perovskite underneath by absorbing high energy photons.

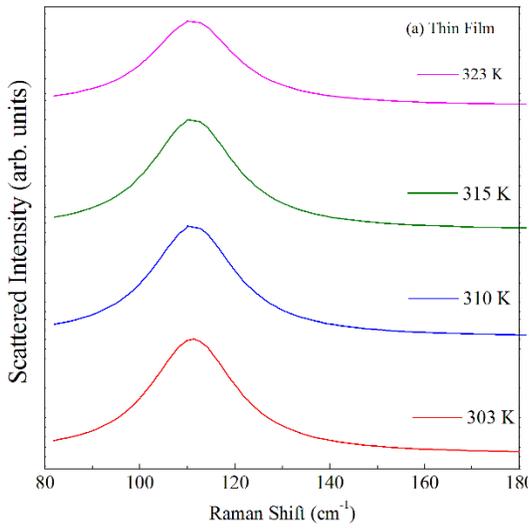
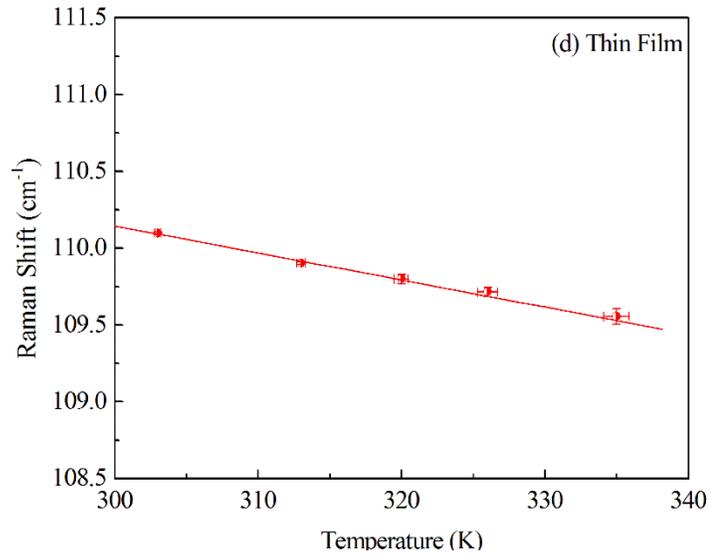
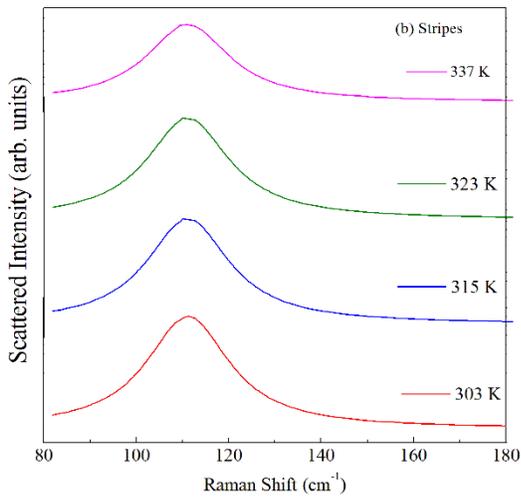
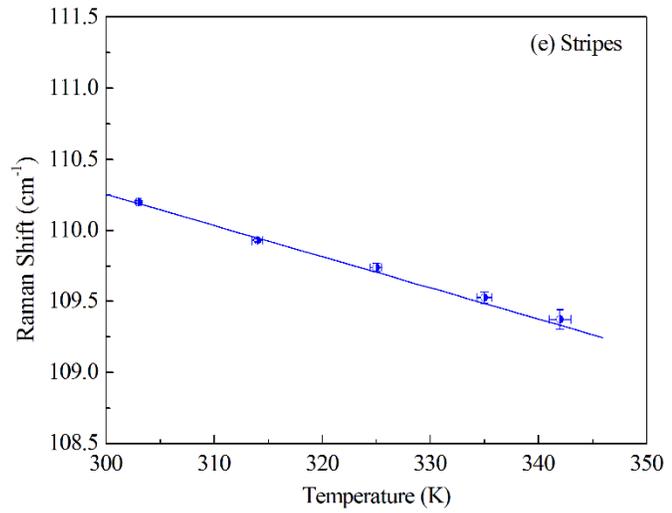
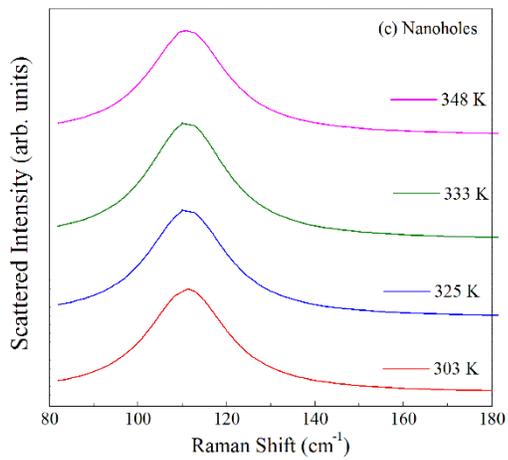
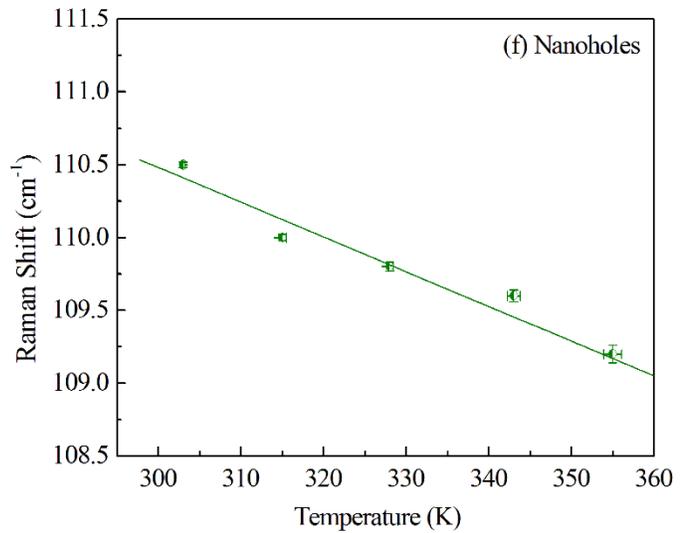

Figure 2: Temperature dependence of the Raman shift for (a), (d) thin film, (b), (e) stripes and (c), (f) nanoholes.

In order to understand the quantitative change in the thermal transport properties we investigate the shift in phonon modes with respect to laser excitation and temperature. Figs 2(d)-(f) show the Raman shift as a function of temperature due to laser induced heating for pristine and patterned perovskite thin films. The Raman frequencies decrease monotonically with temperature. To analyse the red shift in the mode frequencies with temperature we use an equation demonstrating a linear change in Raman frequency with temperature as [64]

$$\omega(T) = \omega_0 + \chi_T T \quad (2)$$

where $\omega_0$ is the phonon frequency at absolute zero, and $\chi_T$ is the first order temperature coefficient. By fitting the phonon downshift using equation (4), the obtained $\chi_T$ values for the pristine, grating and nanohole perovskites are -0.018, -0.020 and -0.022 cm$^{-1}$/K, respectively. It is observed that the perovskite nanoholes exhibit higher $\chi_T$ values than both the pristine and grating perovskites.

It is interesting to observe the differential heat sensitivity of the different textured surfaces on perovskite thin films and therefore can be attractive candidates for different thermal applications including thermoelectric devices. In order to design a thermoelectric device, it is quintessential to have an qualitative estimate of the thermal conductivity of the materials. To get a comparison of the thermal conductivity between normal and patterned perovskites, we have estimated the slope (dω/dP), where P is the laser power, by fitting the data presented in Figs. 4(a)-(c) respectively. The solid lines are the linear fit to the experimental data. The slope for the nanohole (-0.38 cm$^{-1}$/mW) is higher than the striped (-0.32 cm$^{-1}$/mW) and thin film (-0.27 cm$^{-1}$/mW) perovskites respectively. In terms of dω/dP and $\chi_T$, the thermal conductivity (k) can be written as [74]:

$$k = \chi_T \left(\frac{1}{2\pi d}\right)\left(\frac{d\omega}{dP}\right)^{-1} \qquad (3)$$

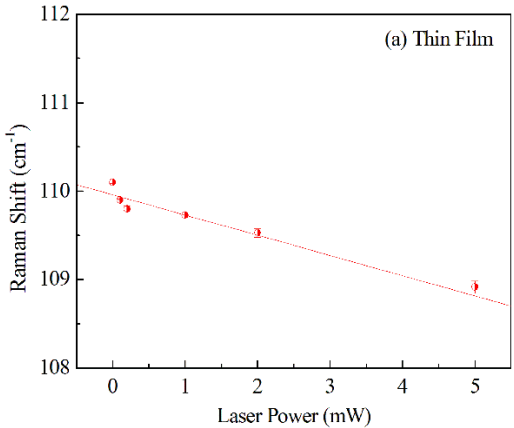

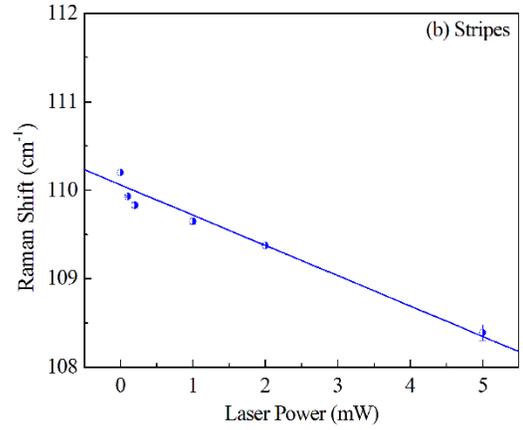

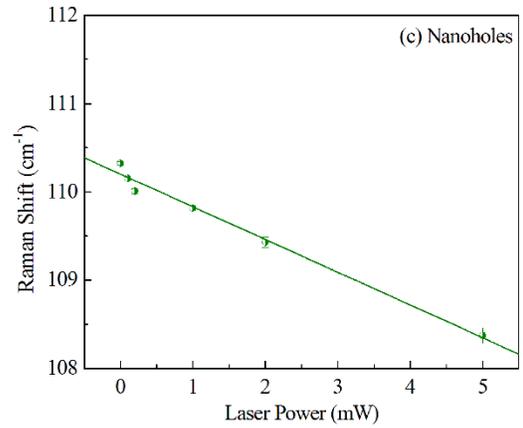

Figure 4: Raman shift of MAPbI thin films as a function of laser power for (a) thin film, (b) stripes and (c) nanoholes.

The nanoholes show a ~11% decrease in the thermal conductivity with respect to the grating structure. Thus, the low thermal conductivity of nanofaceted perovskite thin films compared to the pristine perovskite thin films can make them attractive candidates for thermoelectric applications. Thus, we propose that partial lithographic treatment of perovskite thin film surfaces, with one half planar and the other half lithographically engineered, can be an attractive way to set up thermal gradients required for developing Seebeck potentials necessary for thermoelectric applications. Thus with further optimization of stable textured nanopatterns on thin film perovskite surfaces, a new range of thermal applications can be explored and futuristic technologies can be developed which can act as alternative renewable energy sources.

We have investigated the temperature and power dependent phonon behaviour in spatially patterned MAPbI thin films in order to determine the thermophysical properties of the material. The use of a non-contact and simple spectroscopic tool to study the local heating dynamics at the nanoscale level is useful for a simple and reliable investigation of the thermal phonon behaviour of perovskite thin films. Based on the Raman measurements, the first order temperature coefficient of the phonon shifts has been determined for both pristine and patterned perovskite thin films. The demonstration of altered phonon dynamics in nanohole perovskite thin films as compared to the planar and striped samples reduces the thermal conductivity of the material. Thus, selective nano-patterning of perovskite thin films can serve as an innovative approach to develop localised thermal gradients essential for the fabrication of thermoelectric devices.

†**Corresponding Author**

**Saswata Halder- Department of Applied Physics, Faculty of Science and the Center for Nanoscience and Nanotechnology, The Hebrew University of Jerusalem, Jerusalem, 919041, Israel;**

**Uriel Levy − Department of Applied Physics, Faculty of Science and the Center for Nanoscience and Nanotechnology, The Hebrew University of Jerusalem, Jerusalem 9190401, Israel;**

## ACKNOWLEDGMENT

The authors acknowledge financial support from the Israeli Ministry of Science and Technology. The samples were fabricated, characterized and measured at the Center for Nanoscience and Nanotechnology of the Hebrew University of Jerusalem.

Supplementary Information

for

# Opto-thermal transport engineering in hybrid organic-inorganic lead halide perovskites metasurfaces

Saswata Halder[†], Amit Kessel, Noa Mazurski and Uriel Levy[†]

Department of Applied Physics, Faculty of Science and the Center for Nanoscience and Nanotechnology,

The Hebrew University of Jerusalem, Jerusalem, 91904, Israel.

Corresponding Authors Email: saswata.halder@mail.huji.ac.il; ulevy@mail.huji.ac.il

Results:

## S1. Experimental Details:

S1.1. Synthesis of MAPbI perovskite solution:

Initial ingredients of Methylammonium Iodide (MAI; 99.99% Ossila) and Lead Iodide (PbI$_2$; 99.99%; Sigma Aldrich) were taken in accordance to the stoichiometric ratio. A solution of N, N-dimethylformamide (DMF; 99.99%; Sigma Aldrich) and Dimethyl Sulfoxide (DMSO; 99.99%; Sigma Aldrich) was prepared according to the molar ratio of 4:1. The stoichiometric powder mixture of MAI (0.159 gm) and PbI$_2$ (0.461 gm) were dissolved in the DMF:DMSO solution to prepare a 1M solution (1 ml) inside a nitrogen filled glove box. The solution was kept at 60°C on a hot plate with constant magnetic stirring for 3-5 hours for homogeneous mixing of the powder in the solution.

S1.2. Preparation of poly(dimethyl-siloxane) (PDMS) stamp:

A silicon (Si) chip (20 x 20 mm) was cleaned with "Piranha" ($H_2SO_4:H_2O_2$ 3:1). Two surface imprints of interest were prepared for the present study: (i) A grating with width= 1.3 µm, period= 2.6 µm and height=150 nm, (ii) An array of nanosphere with radius 225 nm. The pattern was defined by spin-coating of an electron beam resist (ZEP 520A) followed by electron beam lithography (ELS-G100 Elionix). The pattern was transferred to the silicon chip using reactive ion etching (RIE) (Corial 200I) with a mixture of $SF_6$ and $CHF_3$. After etching, the remaining ZEP was stripped and the sample was cleaned with "Piranha". A layer of octadecyltrichlorosilane (OTS) was formed on the Si master surface to make the silicon substrate hydrophobic. Gelest h-PDMS base and curing agents were mixed in a ratio of 1:1. The mixture was degassed for a few minutes until all the air bubbles disappeared. The h-PDMS was spin coated on the pre-cleaned Si master for 30 seconds at 2000 revolutions per minute (RPM). The spin coated complex was degassed again to remove any residual air bubbles. The h-PDMS-Si substrate was then partially cured in the oven for 30 min at 65° C, until the h-PDMS felt slightly sticky. SYLGARD 184 base and curing agents were mixed in a ratio of 5:1. The mixture was degassed until all the air bubbles disappeared. s-PDMS was poured on top of the h-PDMS layer and degassed for about 1 hour and subsequently cured at 65° C for 12 hrs. The h-PDMS/s-PDMS stamp was then gently detached from the Si master.

S1.3. Fabrication of patterned thin films:

Glass substrates having dimensions of 15 x 20 mm were cleaned by ultrasonication successively in deionised (DI) water, acetone and isopropyl alcohol and dried with nitrogen gas. They were then further cleaned in an $O_2$ plasma atmosphere for 10 mins to make the surface hydrophilic for uniform wetting of the substrates with the perovskite precursor solution. The cleaned substrates were transferred inside the glovebox and spin coated with the perovskite precursor solution in a two-step process; initially at 1000 revolutions per minute (RPM) for 10 seconds followed by 3000 RPM for 20 seconds. During the spin coating process, toluene solution (15-20 µl) was added dropwise to the precursor solution. This anti-solvent treatment prevents unwanted nucleation of the perovskite solution during spin coat resulting in improved crystallization.[56] For the patterned films, elastomeric PDMS stamp, with the designated patterns, was placed face down on the spin-coated gel and annealed at 90° C under an applied pressure of 4 kgcm$^{-2}$ for 15 minutes for the final crystallization and the removal of excess solvent.

S1.4. Characterization:

The X-ray diffraction (XRD) spectra of the samples were measured using a Bruker D8 Advance X-ray diffractometer in the range of 10°<2θ<80° using a step size of 0.01°. The scanning electron microscope (SEM) images were obtained with an extra high resolution Scanning Electron Microsope Magellan TM 400L. Transmission and reflection spectra were measured using a custom microscope setup. For transmission, the sample was illuminated with a white light source (tungsten-halogen lamp) through a microscope condenser lens. The transmitted light is collected by an objective lens (Nikon, 50×, NA 0.45) and taken to a fiber coupled Ocean Optics Flame spectroscope. Reflection measurements were done in an inverted reflection microscope configuration, using the same light source, objective lens, and spectrometer. The absorption was calculated from reflection and transmission using the relation A=1-R-T. The Raman and photoluminescence (PL) spectra for the samples were measured using a Renishaw inVia Raman Spectrometer which was equipped with a 2400 lines per mm grating and a Peltier cooled CCD. A 50× objective was used to focus the laser beam on the sample with a spot diameter of ∼1 μm. The Raman measurements were carried out at ambient conditions with an excitation wavelength of 514 nm with varying laser power. The temperature at the measurement point was extracted with the help of Eq. 1. The baseline corrected Raman spectrum was deconvoluted with a Lorentzian lineshape and the individual contributions of constituent peaks were obtained by refining the peak position, full width at half maximum (FWHM) and amplitude. The X-ray photoemission spectra of the samples were taken by an X-ray photoelectron spectroscopy (XPS) (Axis Supra). XPS profiles of the samples were acquired using monochromatic Al-K$_\alpha$ source with energy hν = 1486.7 eV and operated at 15 kV and 15 mA. The binding energy was determined with reference to the C 1$s$ line at 284.8 eV.

S2. Soft Lithography of Perovskite Thin Films:

Soft lithography utilizes a soft elastomeric stamp to imprint a desired pattern superficially on a thin film thereby ensuring a conformal contact with flexible substrates and viscous materials. Unlike nano-imprint lithography, soft lithography uses the solution processing route during the synthesis of a perovskite thin film. This allows the crystallization process to occur inside the mold at a mild applied external pressure. Initially, the perovskite solution is deposited on a pre-cleaned glass substrate in a not yet crystallized, viscous thin film. The PDMS mold is then gently pressed on the film, using a customized press followed by heating together to allow for solvent evaporation and crystallization. At the final step, the PDMS stamp is gently removed and an imprinted thin film is obtained. During the imprinting process several competing forces act on the system. The viscous solution is driven into the hollow patterns of the PDMS mold by the applied external pressure as well as capillarity. During the capillary rise, the solvent evaporates and is absorbed in the stamp. In order to achieve a complete filling of the hollow regions of the PDMS mold, the rate of solvent absorption must exceed the rate of evaporation, preventing unwanted accumulation of vapors as well as a corresponding resistive pressure in the gap between the rising perovskite and the mold. The solvent absorption rate is constant and depends on the PDMS properties. The rate of evaporation on the other hand depends on the solvent properties [1] and annealing temperature. Temperature plays an essential role as the rate of solvent evaporation is a thermally dependent phenomena which can lead to crystallization even before complete filling the hollow regions in the stamp. In the case of MAPbI, standard annealing temperatures (90–100 °C) as well as an external applied pressure ≈4 kgcm$^{-2}$ is sufficient for a complete transfer of the pattern. We observed that further increasing the force does not improve the quality of the imprint.

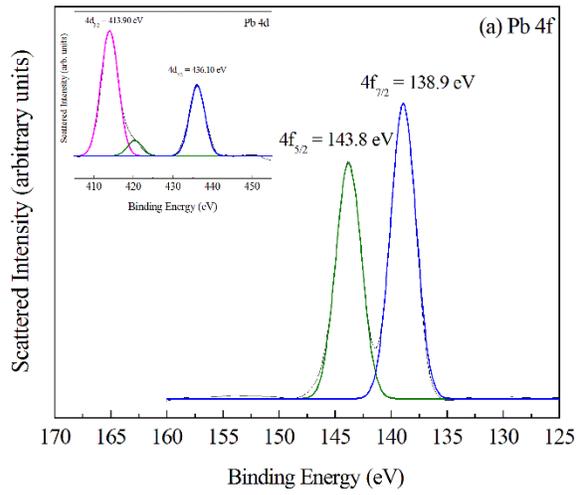
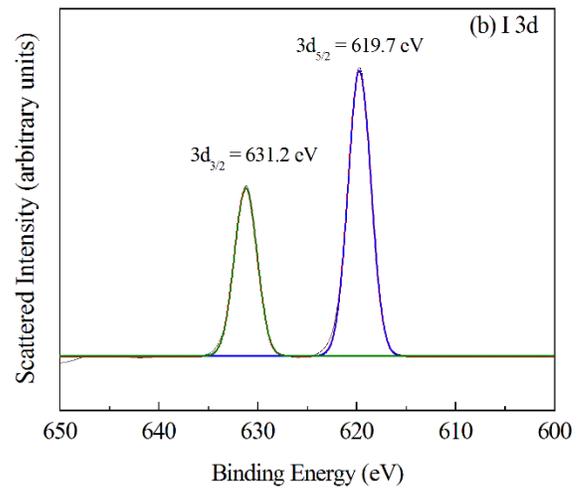
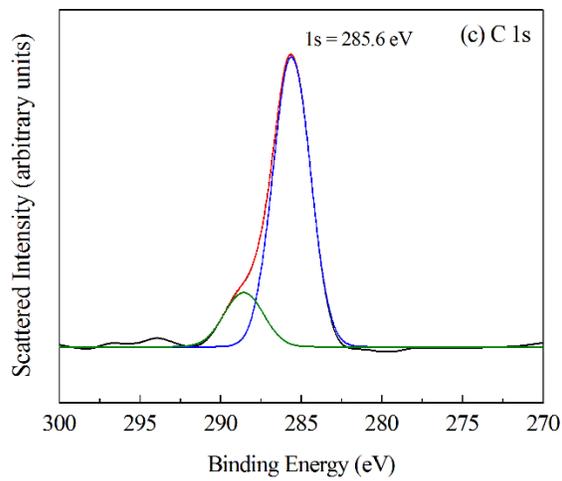
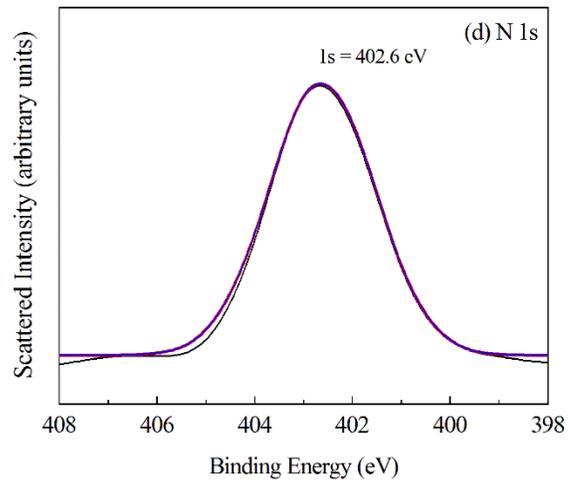

**Figure S1**: XPS core level spectra of pristine perovskite thin films: Pb-4f (a), I-3d (b), C-1s (c) and N-1s (d) spectra of pristine perovskite thin films

The core-level XPS spectra for MAPbI is shown in Fig. S2(a)-(d). The XPS spectra displays the characteristic signature peaks of MAPbI. The Pb 4f spectrum exhibits peaks at 138.9 eV and 143.8 eV corresponding to the Pb $4f_{7/2}$ and Pb $4f_{5/2}$ state associated to the $Pb^{2+}$ state in $PbI_3^-$ octahedra. The C-1s peak shows two peaks corresponding to the presence of C-H/C-C and C-N bonds. The I-3d orbital shows the presence of two peaks at 619.7 eV and 631.2 eV corresponding to $I-3d_{5/2}$ and $I-3d_{3/2}$ states respectively.

S3: Microstructure:

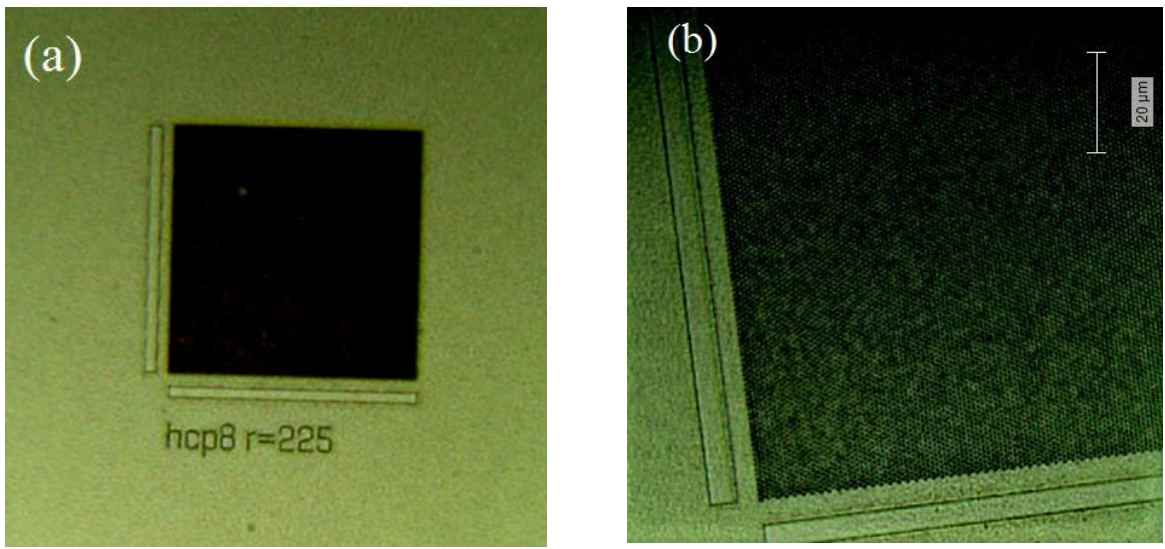

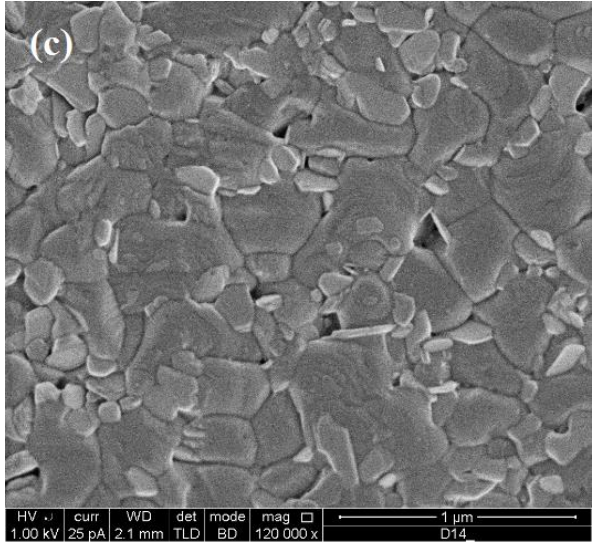

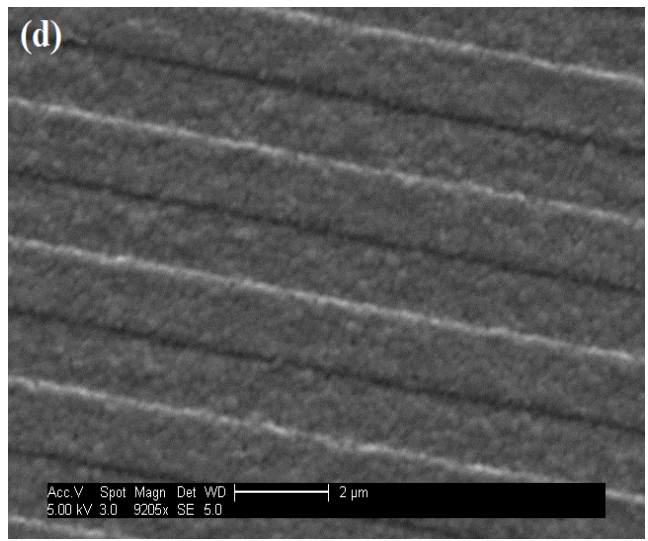

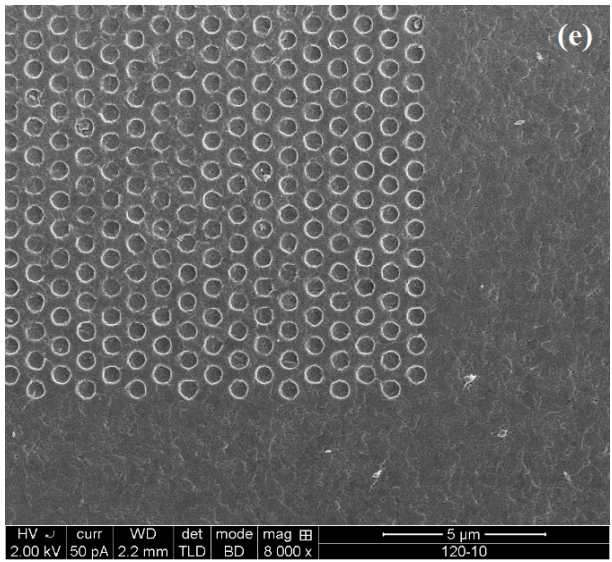

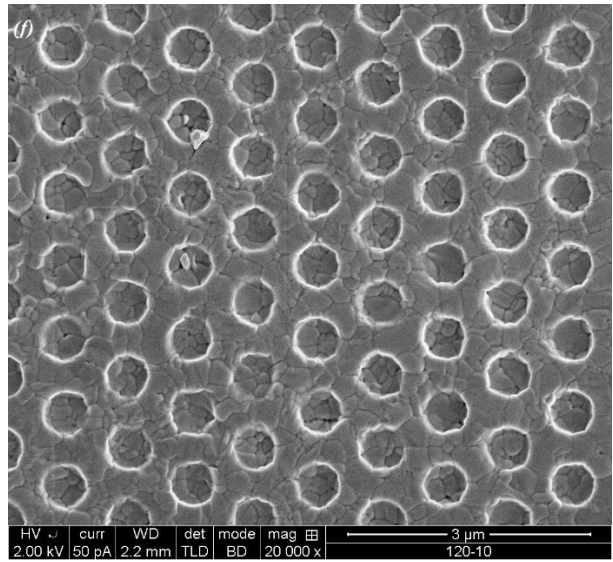

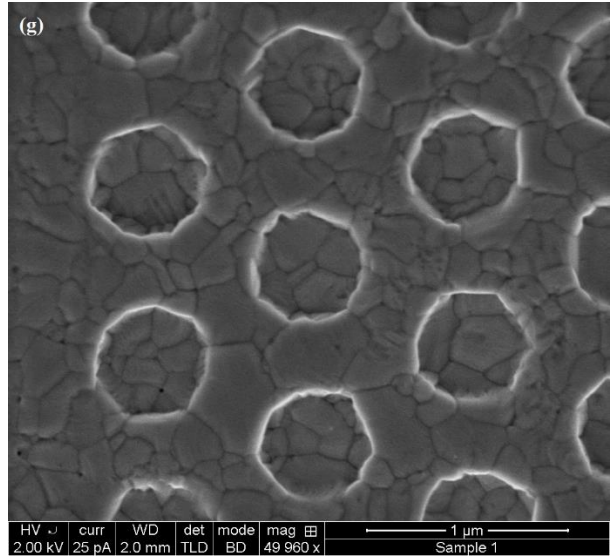

**Figure S2**: (a)-(b) Optical microscope images for nanoholes of radius 225 nm. The higher contrast in the patterned region with respect to the outer planar region of the perovskite signifies higher absorption. (c) SEM images of the bare unpatterned perovskite with voids signifying an average crystallization process. (d) SEM image of MAPbI thin film partially treated using a flat PDMS stamp and a patterned PDMS stamp. The use of the PDMS stamp clearly enhances the crystalline nature of the samples as seen from the reduction in the voids in the lithographically treated samples with respect to the untreated sample. (e)-(f) SEM images of perovskite metasurfaces with stripes and holes with dense arrangement of grains.

S4. Raman Spectroscopy:

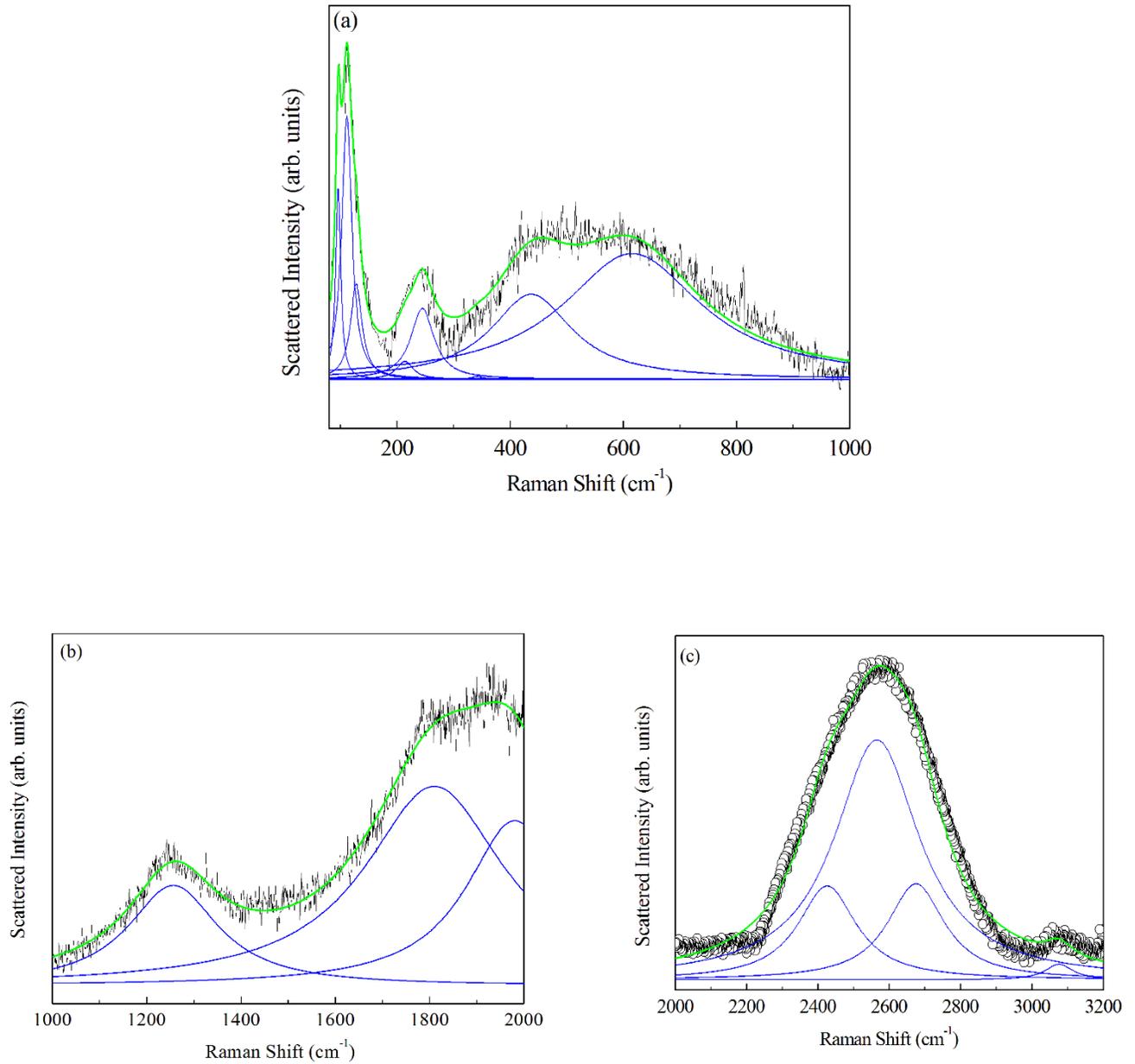

**Figure S3**: (a)-(c) Deconvoluted Raman profiles at different intervals. The blue curves represent the individual contributions to the Raman spectrum. The green curves represent

the summation of the blue curves in each spectral regime. The black curves represent the measured results.

The structure of the Raman spectrum at lower wavenumbers (80–200 cm$^{-1}$) consists of a number of major contributions at 96, 110 and 126 cm$^{-1}$. The spectrum at higher wavenumbers regime is dominated by the contributions from 214, 250, 340, 448 and 620 cm$^{-1}$. From a qualitative group theoretical analysis, the number of eigenmodes in this wavenumber domain should be 18, of which three are acoustic vibrations and the rest 15 are optical modes. However, due to the low strain in the system, most of the modes are energetically degenerate and therefore their individual experimental resolution is difficult. In the perovskite structure, the organic MA cation undergoes different reorientations about its lattice position besides coupling to the inorganic lead-halide framework, thereby further complicating the overall dynamical picture. The low wavenumber features are associated with cage dominated optical modes consisting of the overall translation and rotation of the organic cation coupled to the lattice motions. Inspection of the normal modes calculated in the region around 96 cm$^{-1}$ reveals a major component from librations of the organic cations and from Pb−I stretching. Thus, the 96 cm$^{-1}$ band is reasonably associated with both the Pb−I stretching and mainly provides information regarding the inorganic component of the material. The Raman band at 110 cm$^{-1}$ is associated with the motion of the organic cations.[3] A similar result holds for the weak band measured at 160 cm$^{-1}$, that can be safely associated with the band calculated at 156 cm$^{-1}$, corresponding to the libration of the organic cations [4].

The region between 100 and 200 cm$^{-1}$ is associated to librational and molecular motions of the MA cations due to the mismatch in the mass of the organic and inorganic cation.

Density functional theory (DFT) based electronic structure calculations elucidate how the interactions between the organic cations and the inorganic counterpart may affect the vibrational frequency and the Raman intensity of the MA torsional mode in MAPbI. [3] The broad and unresolved band at 200−340 cm$^{-1}$ is assigned to the torsional mode of the MA cations. The frequency of the MA torsional modes in the crystal structures is a result of two competing effects, associated with the interactions between the MA cation and the inorganic cage. On the one hand, the deformation of the MA molecule shifts the torsional mode toward higher frequencies; on the other hand, the formation of specific hydrogen bonding interactions, typical of these compounds shifts this mode towards lower frequencies. Due to this disorder of the organic cations in MAPbI, the MA cations are subjected to different balances between molecular deformation and hydrogen bonding interactions. This effect spreads the torsional frequencies over a broad range of values. This mode therefore acts as the possible marker for the orientational ordering of the MA cations in the material. The higher wavenumber peaks in the range 400-3000 cm$^{-1}$ result from the stretching/breathing motions of the MA cations alone, namely, $CH_3$ asymmetric breathing, $NH_3$ asymmetric breathing, C-H symmetric stretching and N-H symmetric stretching respectively.[3] These peaks tend to remain quasiharmonic under temperature change showing that the associated vibrations are ideally not harmonic but also do not couple to other vibrations through their anharmonic terms.

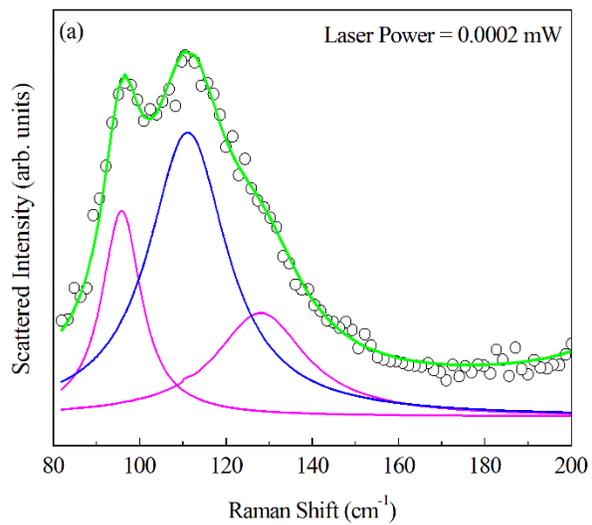
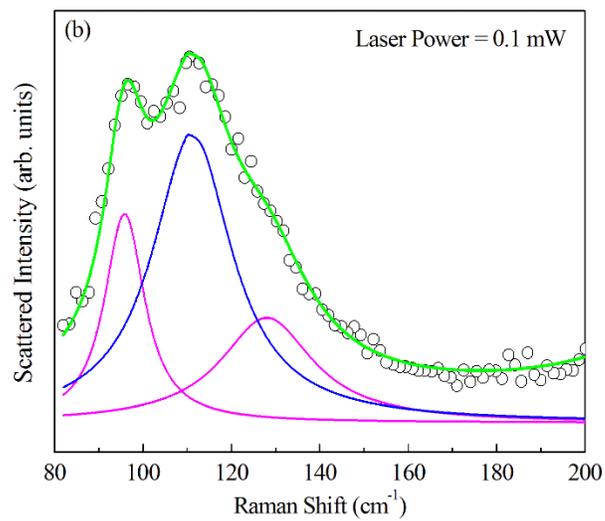
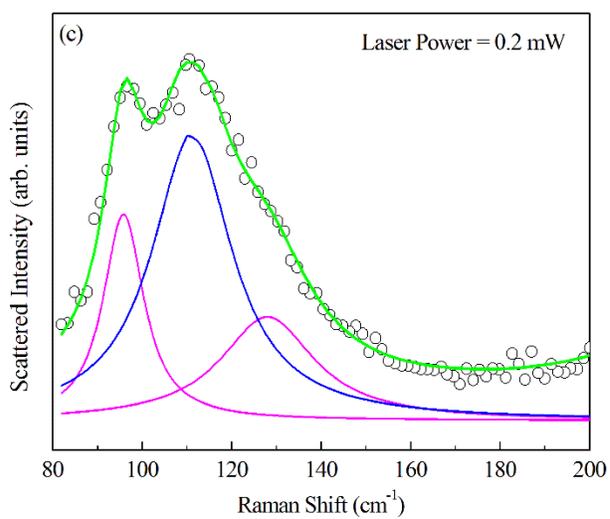
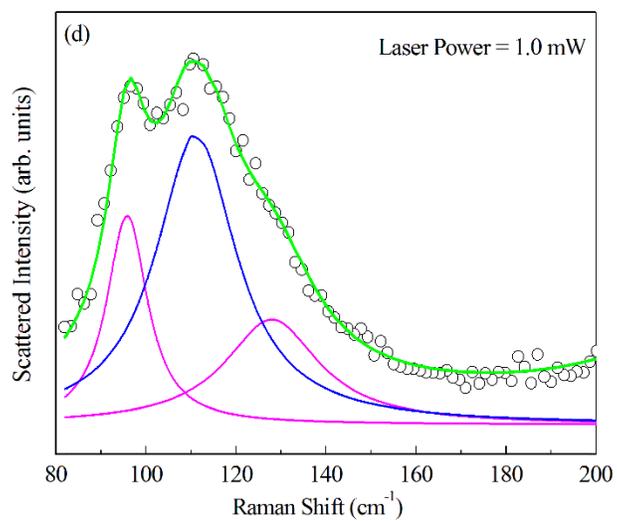
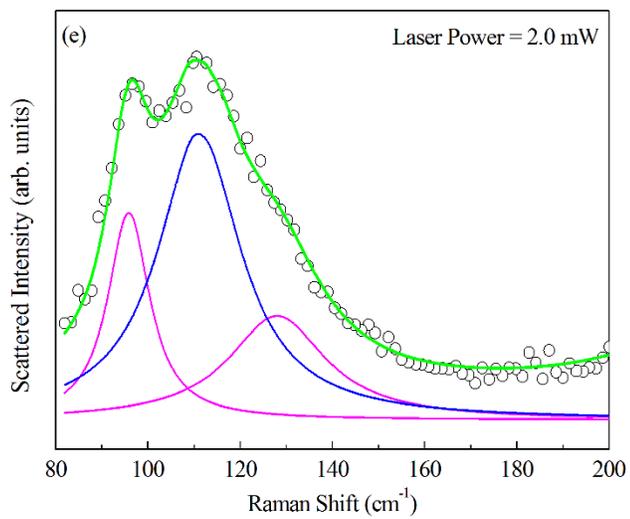

**Figure S4**: (a)-(e) Deconvoluted Raman profiles for MAPbI planar films at different laser powers. The blue deconvoluted curve is the Raman mode of interest in this study.

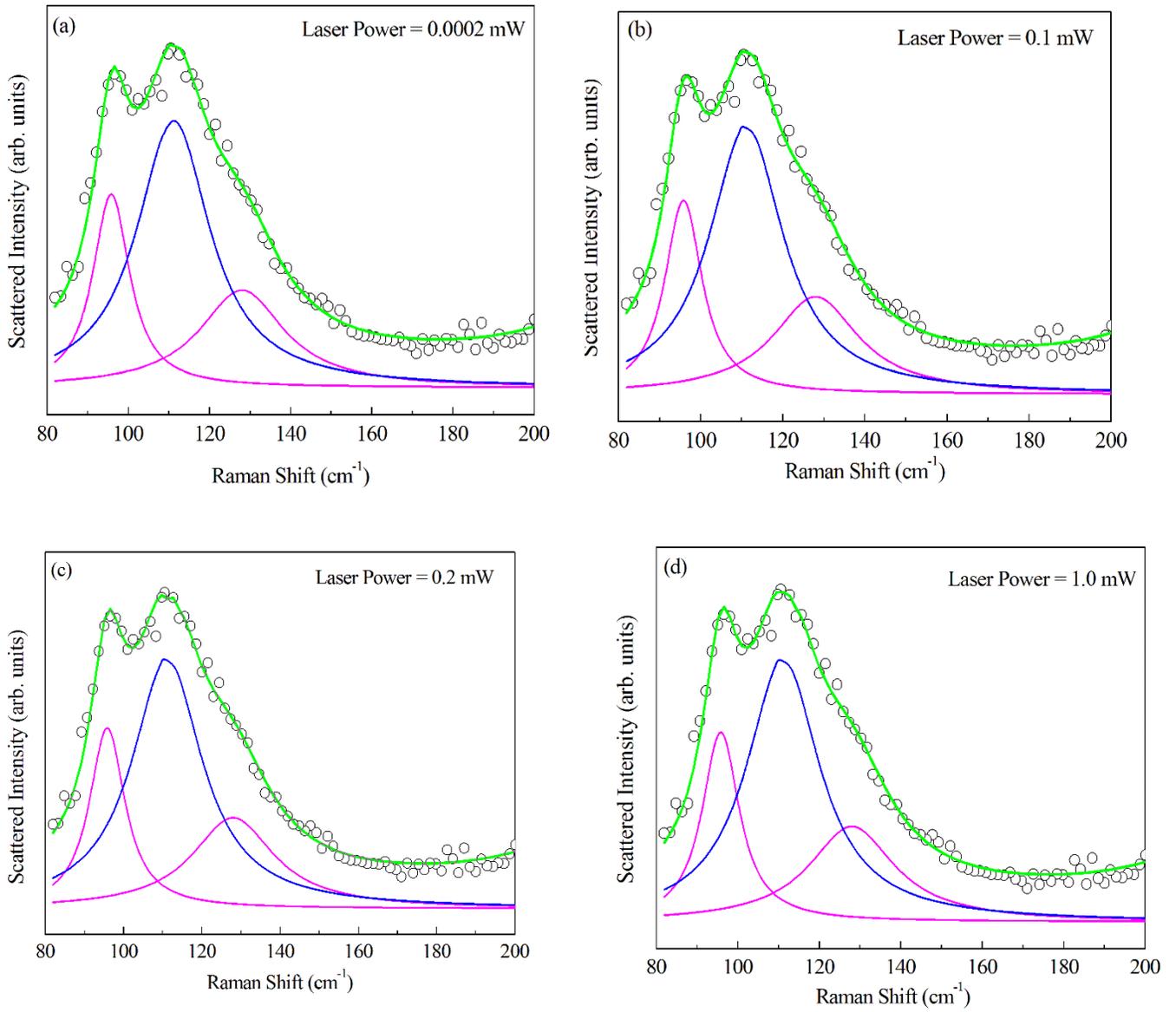

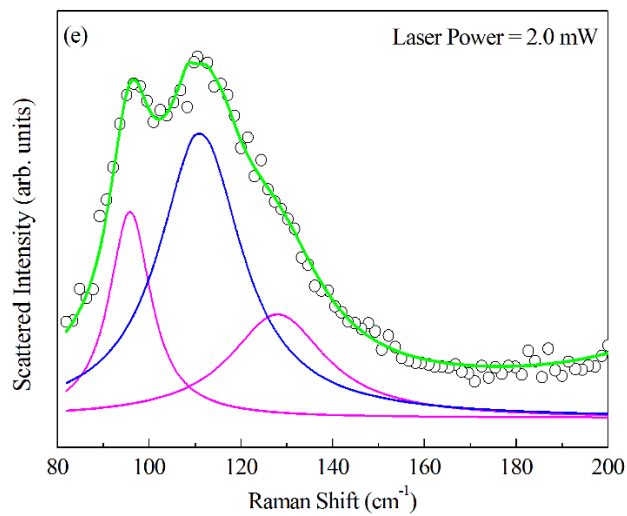

**Figure S5**: (a)-(e) Deconvoluted Raman profiles for striped MAPbI films at different laser powers. The blue deconvoluted curve is the Raman mode of interest in this study.

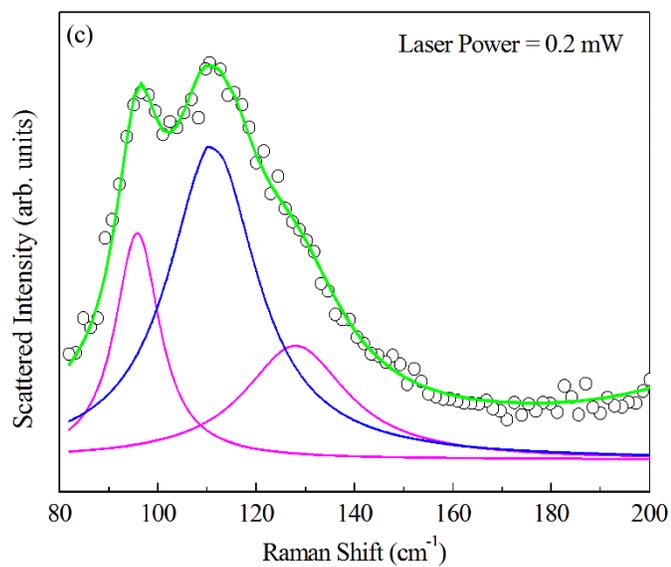

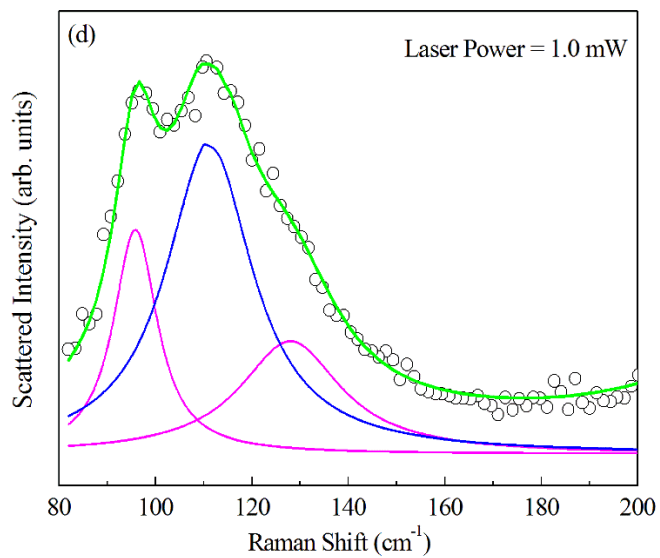

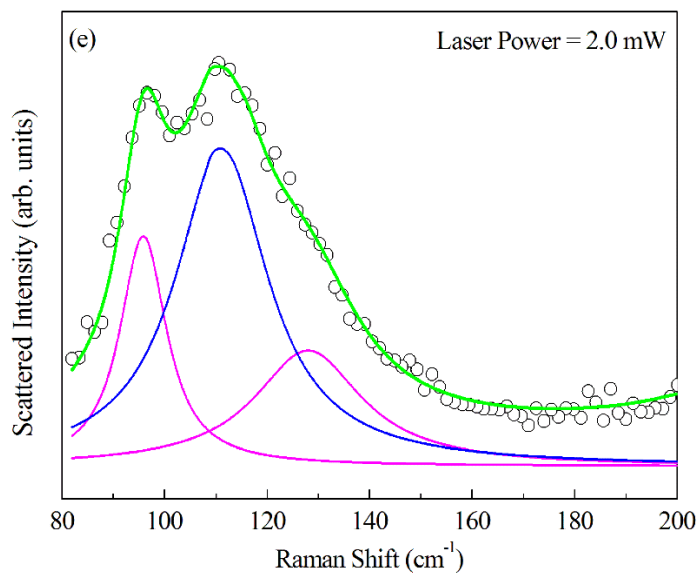

**Figure S6**: (a)-(e) Deconvoluted Raman profiles for MAPbI nanoholes at different laser powers. The blue deconvoluted curve is the Raman mode of interest in this study.